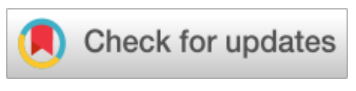

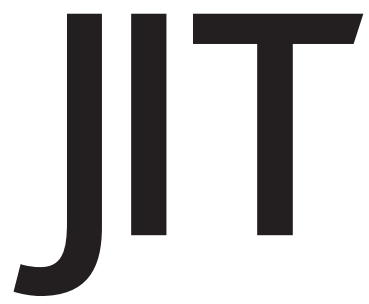

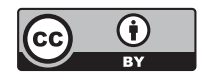

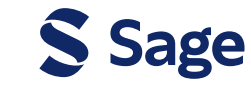

# Generative artificial intelligence for literature reviews

**Gerit Wagner[1], Julian Prester[2], Reza Mousavi[3], Roman Lukyanenko[3] and Guy Paré[4]**

**Abstract**

Generative artificial intelligence (GenAI), based on large-language models (LLMs), such as ChatGPT, has taken organizations, academia, and the public by storm. In particular, impressive GenAI capabilities such as summarization of large text corpora, question-answering, data extraction, and translation, carry profound implications for the conduct of literature reviews. This impacts science, organizations and the general public, as all can benefit from GenAI-supported literature reviews. Building on the technical foundations of GenAI and grounded in established methodological discourse, this work outlines approaches for conducting literature reviews using both general-purpose (e.g., ChatGPT, Gemini, Claude) and specialized GenAI tools (e.g., Consensus, Elicit). We provide illustrative examples of prompts and suggest methodologically-sound literature review strategies. Throughout this perspective paper, we adopt a balanced approach considering both the opportunities and the risks of relying on GenAI in the conduct of literature reviews. We conclude by discussing philosophical questions related to the effects of GenAI on long-term scientific progress, and also present fruitful opportunities for research on improving the core of GenAI's technology—its architecture and training data—and suggest open issues in GenAI-based literature reviews methodology.

## Introduction

Generative artificial intelligence (GenAI), particularly in the form of large language models (LLMs) such as ChatGPT and Gemini, has rapidly gained popularity in organizations, academia, and among the general public. It is widely viewed as a transformative development, especially for knowledge-intensive tasks involving language, including summarization, question answering, and synthesis. At the same time, assessments of GenAI's impact remain contested. While some users report substantial gains in efficiency and convenience, others point to disappointing performance, limited real-world value realization, and recurring cycles of hype and disillusionment associated with earlier waves of AI technology.

Notwithstanding these divergent views, the pace of recent advances in GenAI has been striking, raising fundamental questions about how such systems may reshape established scientific practices. Although GenAI holds considerable promise for advancing research, it also presents significant challenges. The exponential growth of scientific publications already imposes a substantial cognitive load on researchers, increasing the likelihood that they overlook relevant and timely findings (Bornmann et al., 2021; Thelwall and Pardeep, 2022). In this context, GenAI introduces a profound paradox: while these technologies may further accelerate the production of scholarly content, thereby intensifying informational overload, they also offer powerful new capabilities for synthesizing large bodies of literature and mitigating the very complexity they help create.

Within the specific domain of literature reviews, Generative artificial intelligence can support a wide range of

[1]Department of Information Systems and Applied Computer Sciences, University of Bamberg, Bamberg, Germany
[2]School of Strategy, Innovation and Technology, University of Sydney, Sydney, NSW, Australia
[3]McIntire School of Commerce, University of Virginia, Charlottesville, VA, USA
[4]Department of Information Technologies, HEC Montréal, Montreal, QC, Canada

**Corresponding author:**
Guy Paré, Department of Information Technologies, HEC Montréal, 3000, chemin de la Côte-Sainte-Catherine, Montreal, QC, H3T 2A7, Canada.
Email: guy.pare@hec.ca

activities, from foundational tasks such as exploratory searching and summarization to more complex functions involving project management and conceptual knowledge synthesis (Alavi et al., 2024; Schryen et al., 2024; Susarla et al., 2023). However, not all observers are convinced of their unqualified benefits. Some caution that an overreliance on GenAI may erode core scientific skills by discouraging deep engagement with primary sources (Zur Schlemmer, 2024). Accordingly, while GenAI may facilitate more comprehensive and efficient evidence synthesis, its role in scientific inquiry warrants careful and balanced consideration of both its potential benefits and its associated risks.

This need for caution is reinforced by the considerable uncertainty surrounding GenAI's ongoing development. It thus remains unclear how GenAI will ultimately reshape research, as the broader GenAI ecosystem—including transformer architectures, pre-training pipelines, alignment and safety protocols, and the applications built on them—continues to evolve rapidly. Compounding this uncertainty is the continual discovery of emergent properties and unforeseen functionalities within GenAI—a phenomenon whereby novel, qualitatively distinct capabilities, such as multi-step reasoning, manifest only after models surpass a critical threshold of scale (Wei et al., 2022a, 2022c). Consequently, many advanced capabilities are not the product of targeted engineering but are instead outcomes of revised scaling principles and empirical discovery (Hoffmann et al., 2022; Kaddour et al., 2023).

This discovery-oriented paradigm poses fundamental challenges to interpretability, leaving significant gaps in understanding model mechanisms and in reliably steering their behavior (Bowman, 2023). The continued adaptation of these systems to novel problems and domains further underscores how little is known about the ultimate boundaries of their capabilities. Unsurprisingly, experts hold widely diverging expectations for the future of GenAI, ranging from its containment within narrow, regulated use cases to the eventual emergence of artificial general intelligence (Bubeck et al., 2023; Hubert et al., 2024).

Considering the existing uncertainty related to the future of GenAI, we believe it is instructive to discuss possible opportunities, modalities, and risks related to the use of GenAI in the conduct of literature reviews. While the initial discourse has quickly produced suggestions on how GenAI could be of use in our context (e.g., Alshami et al., 2023; Rahman et al., 2023; Temsah et al., 2023), these preprints, commentaries, and studies do not offer a substantial connection to the established methodological knowledge. Without considering how GenAI-enhanced literature reviews can be reconciled with the goals and types of reviews (Paré et al., 2015), the activities of the process, systematicity of methodological choices, as well as reporting requirements (Paré et al., 2016; Templier and Paré 2018), the use of GenAI for literature reviews may struggle to produce reviews of good quality and fail to meet expectations of an accepted and valid review process. Instead, it remains essential that researchers are knowledgeable in their domain, understand the nuances of literature review methods, and leverage GenAI considerately (Qureshi et al., 2023). In doing so, we believe it is important to discuss how established methodological practices should be continued and how GenAI can enhance the conduct of literature reviews.

The primary goal of this paper is to discuss how GenAI transforms the conduct of literature reviews, provide constructive suggestions for prospective authors, and discuss potential risks and opportunities. It continues our work on the use of previous generations of AI[1] for literature reviews (Wagner et al., 2022) and discusses how recent advances in GenAI could affect literature review practices in the future.

This paper is relevant for literature reviews across a wide range of scientific disciplines and research genres. It is also written for a broader audience, as organizations, journalists and the public increasingly use GenAI to conduct reviews of their own. At the same time, we deliberately choose a specific context for our examples. Our running examples are GenAI-supported reviews in the context of information systems design and use. First, information systems design and use is the context we are most familiar with. We have undertaken numerous manual reviews and reviews with the support of previous generations of AI tools, examining various aspects of information systems design and use (Dissanayake et al., 2025; Larsen et al., 2025; Prester et al., 2021; Recker et al., 2021). We are well positioned to interpret the performance and findings of GenAI tools in this context.

Furthermore, information systems, as a proximal discipline to GenAI, is already actively involved in understanding the use of GenAI (e.g., Alavi et al., 2024; Ngwenyama and Rowe, 2024; Storey et al., 2025; Susarla et al., 2023). Finally, the information systems discipline has developed a vibrant methodological discourse on literature reviews, including on AI-supported reviews (e.g., Boell and Cecez-Kecmanovic, 2015; Paré et al., 2024; Storey et al., 2025; Templier and Paré, 2018; Wagner et al., 2022), increasingly serving as a reference to other disciplines (e.g., Aguinis et al., 2023).

The remainder of this paper is structured as follows. First, we establish a conceptual foundation in GenAI, distinguishing it from traditional AI, and surveying its primary modalities. We also introduce effective prompting strategies as the core method for interacting with these systems. Next, we build on these concepts to offer suggestions for the use of GenAI and corresponding prompts for the different activities of the review process. Finally, and before concluding the paper, we discuss the broader opportunities, challenges and open questions related to the use of GenAI in the conduct of literature reviews.

## Technological foundations

### GenAI versus AI

GenAI represents a significant evolution from traditional AI, shifting the technological paradigm from data analysis to content creation. Traditional AI approaches are primarily discriminative; they excel at pattern recognition, classification, and predictive analytics based on existing datasets. Their main purpose is to interpret and make judgments about patterns in data. In contrast, GenAI models are different: their core function is to produce new, synthetic content that mimics the patterns and structures of the data on which they were trained. This capability spans a wide array of modalities, allowing these systems to compose text, create realistic images, write computer code, synthesize audio, and even design molecular structures (Brown et al., 2020; Zhong et al., 2024). This marks a fundamental shift from technologies that primarily analyze existing information to those that can synthesize novel artifacts.

This distinction is profound in the context of research. While a traditional model might classify scholarly articles or predict trends, GenAI can actively participate in the research process. For instance, it can assist in problem formulation by drafting hypotheses (text) or creating conceptual diagrams (images). It can enhance data analysis by generating code to process datasets or by creating synthetic data for model validation. For dissemination, it can draft manuscript sections (text), design figures (images), or even compose a score for a video abstract (audio). This shift from merely automating repetitive tasks to actively contributing creative input and generating new content underscores the transformative potential of GenAI in redefining the conduct of AI-supported literature reviews (AILRs).

## *GenAI modalities*

The diverse applications outlined above are enabled by distinct classes of generative models, each specialized for a specific data modality. The primary modalities include text, image, audio, video, and integrated multimodal systems.

The text modality is foundational to GenAI, powered by the immense capabilities of LLMs such as GPT-4 and Llama 3.1. The development of these models marks a pivotal departure from earlier deep learning approaches in natural language processing (NLP). For years, the field was dominated by sequential architectures like recurrent neural networks (RNNs) and their more advanced variant, long short-term memory (LSTMs). These models processed text one word at a time, maintaining a "memory" of prior context. However, this sequential method faced two critical limitations: first, it struggled to maintain context over long passages, as information would degrade or "vanish" across many steps (Zhao et al., 2020); second, its word-by-word nature prevented the parallelization needed to train on internet-scale datasets (Hwang and Sung, 2015).

The introduction of the Transformer architecture in 2017 was a paradigm shift that solved these problems (Vaswani et al., 2017). Its key innovation, the self-attention mechanism, abandoned sequential processing entirely. Instead, it allowed the model to weigh the influence of all words in a sequence simultaneously, creating direct pathways for context to flow regardless of distance and thus capturing complex, long-range dependencies.

Crucially, this architecture's design was highly parallelizable, making it computationally feasible to train models of unprecedented size (Devlin et al., 2019). This scalability is what directly paved the way for modern LLMs (Radford et al., 2018). By dramatically increasing model parameters and training data, researchers discovered that scaled-up Transformer models exhibited the sophisticated, emergent capabilities that have since catalyzed a revolution across NLP, enabling applications from nuanced conversational agents to complex code generation (Brown et al., 2020).

Concurrently with the revolution in text generation, a separate lineage of architectural innovation was enabling the synthesis of rich media. Beginning with generative adversarial networks (GANs) (Goodfellow et al., 2014) and later advancing significantly with diffusion models (Ho et al., 2020; Sohl-Dickstein et al., 2015), these techniques resulted in models that generate high-fidelity images, audio, and video from descriptive text prompts. This progress is exemplified by tools like DALL·E[2] for image generation and Sora[3] for video generation, which can translate linguistic concepts into detailed visual content.

The current frontier of GenAI is defined by the convergence of two powerful streams: the linguistic prowess of Transformer-based LLMs and the sensory generation capabilities of architectures like diffusion models. The evolution toward today's multimodal systems began with foundational techniques designed to bridge the gap between different data types. A pioneering step in this direction was the development of joint embedding spaces, exemplified by models like CLIP (Contrastive Language–Image Pretraining) (Radford et al., 2021). By learning to align text and images within a shared representational framework, CLIP enabled models to achieve a cross-modal understanding for tasks like zero-shot image classification and text-based image retrieval, laying the essential groundwork for more complex integrations.

The evolution from these initial integrations to today's state-of-the-art models has been rapid, marked by a fundamental shift in architectural philosophy. Whereas early multimodal applications often relied on a pipeline of separate, specialized models (e.g., a speech-to-text model feeding into an LLM, which then outputs to a text-to-speech model), the current frontier is defined by single, unified models trained end-to-end. This paradigm shift is exemplified by Google's Gemini, which was designed to be "natively multimodal" from its inception, capable of reasoning seamlessly across text, images, video, and audio within one cohesive architecture (Gemini Team Google et al., 2023), resulting in tools such as NotebookLM.[4] Similarly, OpenAI's GPT-4o ("o" for "omni")[5] replaced its prior model pipeline with a unified system, drastically reducing latency and enabling fluid, real-time interaction across text, audio, and images. This architectural leap allows current models to perform highly complex cross-modal tasks, such as answering verbal questions about a live video feed or interpreting emotional tone from audio, within a single, coherent system.

While the ability of these state-of-the-art models to process text, audio, and video represents a transformative frontier for many research fields, their application to the literature review process hinges primarily on their sophisticated textual capabilities. Scholarly knowledge is overwhelmingly codified and disseminated through text, making the core activities of a literature review—from source identification to synthesis and narrative construction—fundamentally text-centric endeavors.

Accordingly, our analysis focuses on GenAI models renowned for their robust text processing. This includes both LLMs like GPT-4 and Llama 3.1,[6] and leading multimodal systems such as GPT-4o, Gemini, and Claude 3.5 Sonnet, whose underlying linguistic engines are paramount for this work.[7] Harnessing the power of these models for academic purposes requires deliberate interaction strategies (though we do explore opportunities for non-textual formats in a later section). The subsequent section, therefore, delves into methodologies for interfacing with GenAI, emphasizing prompting techniques that optimize its effectiveness in research settings.

## *Prompting strategies for GenAI*

Given the intrinsic dependence of LLMs and multimodal GenAI on the initial prompt for generating text, the selection of an appropriate prompt is critical. A spectrum of prompting strategies exists, each tailored to enhance the model's performance in specific contexts. In the subsequent discussion, we delve into various prompting strategies that hold particular relevance for conducting literature reviews. These strategies are designed not only to refine the model's output in terms of relevance and specificity but also to ensure that the synthesized reviews are comprehensive, accurately reflecting the breadth and depth of the existing scholarly discourse. This approach underscores the necessity of strategic prompt design as a fundamental step in leveraging GenAI for academic and research purposes, particularly in the meticulous task of a literature review. When applying GenAI to literature reviews in academic research, several prompting strategies stand out for their effectiveness:

1. Exploratory prompting: This strategy involves asking open-ended questions to explore broad themes or identify under-researched areas within a field (Sun and Wang, 2025). This approach mirrors exploratory search and information-seeking behaviors that researchers employ when navigating unfamiliar domains or scoping new research directions (Gusenbauer and Haddaway, 2021). For example, it is particularly effective in the exploratory stages of a literature review, where the goal is to map out the landscape of existing research. An example of the exploratory prompting strategy applied to the literature review process is provided in problem formulation prompts (Tables 1 and 3).
2. Zero-shot and few-shot learning (Touvron et al., 2023): These techniques are particularly useful when dealing with highly specialized or emerging topics in research, where pre-existing examples or detailed training data may be limited. For example, by providing the model with a definition of a task, such as the formulation of a search strategy, and possibly a few examples (few-shot) or none at all (zero-shot), researchers can prompt GenAI to generate insights or identify trends in a literature corpus that have not been explicitly programmed into its training data. An example of the few-shot prompting strategy applied to the literature review process is provided in the search query prompt (Table 5).
3. Chain-of-thought prompting (Wei et al., 2022b): This strategy involves guiding the GenAI model through a logical reasoning process, breaking down complex queries into simpler, sequential steps. For example, when exploring the impact of digital transformation on organizational culture, a researcher might use a chain of thought prompting to first identify key components of digital transformation, then assess their influences on different aspects of organizational culture, and finally synthesize these impacts into a coherent narrative. This step-by-step approach helps in structuring the literature review process and ensures that the model's outputs are not only relevant but also logically sound. An example of the chain-of-thought prompting strategy applied to the literature review process is provided in the data extraction prompt or the data analysis and synthesis prompt (Table 9).
4. Retrieval-augmented generation (RAG) (Lewis et al., 2020): RAG is a technique that enhances LLMs by combining the pre-trained LLM model with additional sources provided by the user before generating a response. This makes the GenAI responses more context-aware, relevant to the specific task, and less prone to "hallucinations." Although RAG is not strictly a prompting strategy, it integrates the generative capabilities of models with the strength of information retrieval, enhancing the quality and relevance of responses. This makes it particularly valuable for complex tasks like literature reviews. For example, in the context of reviewing literature on the impact of digital transformation on organizational culture, a RAG prompt can leverage both the retrieval of existing scholarly articles and the generative aspect to synthesize and analyze findings. An example of the RAG prompting strategy applied to the literature review process is provided in the first literature search prompt (Table 4).

In addition to these strategies, role-based or persona prompting can enhance output quality by prefixing prompts with statements such as "You are an expert in…." This technique establishes the model's context and expertise level, activating domain-relevant knowledge and improving response quality (Salewski et al., 2023). Several prompts in this paper employ this technique to guide the model toward more specialized outputs.

In sum, by leveraging these prompting strategies properly and responsibly, researchers can harness the capabilities of GenAI to conduct more efficient, thorough, and insightful literature reviews. These approaches not only save time but also enhance the depth and breadth of the review process, enabling scholars to uncover novel insights and contribute more meaningfully to their fields of study.

## Applications of GenAI in the literature review process

We now provide the most prevalent opportunities for applying GenAI in the conduct of literature reviews, to sensitize readers to methodological nuances when using GenAI, and to anticipate how GenAI may change the review process in the future. To accomplish this, we adopt an iterative conception of the literature review process, in which researchers select and revisit literature review activities without following a strict, predefined sequence (cf. Boell and Cecez-Kecmanovic, 2014). For each review activity, we provide example prompts and short evaluations based on the authors' assessment of prompt outputs.

In discussing GenAI capabilities, we refer to HuggingFace,[8] an online platform that provides a toolkit library, open-source LLMs, an active community, and educational resources for deep learning-based models, and HELM,[9] which provides an overview of state-of-the-art LLM evaluation results across a variety of tasks. While models covered in HuggingFace can be applied to different types of data (including audio, photos, and video), the NLP and multimodal capabilities are particularly interesting for our purposes. They include text generation, question answering, summarization, translation, document question answering, image-text-to-text, and any-to-any format generation.

There are three key premises for our work. First, researchers must be familiar with common methodological practices, such as goals of reviews (Paré et al., 2015, 2024; Rowe, 2014), choices in different steps (Templier and Paré, 2018), as well as transparency and systematicity requirements (Paré et al., 2016). At least for now, GenAI needs explicit prompts and context to provide adequate output. In the future, custom GPT versions may combine instructions with extra knowledge and any combination of skills. Second, we expect researchers to ensure that GenAI tools have access to relevant full-text documents, typically by downloading PDFs and providing them with the prompts. This aligns with the reporting requirements for standalone reviews, which ask authors to control, track and report on the retrieved, screened, and analyzed papers (Templier and Paré, 2018). It is important to note that GenAI tools offered by individual publishers, or operating on open-access papers without considering the specific sample of the review project, are more suitable for informal reviews or exploratory activities, rather than the standalone review process. Third, researchers must be aware that entering the prompts does not guarantee a useful outcome for all review projects and that adequate oversight is mandatory. This means that all results provided by GenAI must be fact-checked, evaluated critically, and disclosed appropriately. As Tingelhoff et al. (2025) put it, researchers must carefully evaluate "what we should allow Gen.AI to do" (p. 78). In addition, it needs to be checked whether full-text documents (PDFs) provided with the prompt fit into the review scope, or whether task-splitting or paid API-access is needed. To support these activities, corresponding research software will need to offer new functionality related to transparent versioning and validation of literature review data.

### *Problem formulation*

When embarking on a standalone review paper, the problem formulation involves identifying a promising opportunity, assessing the feasibility of the project, and preparing the groundwork for the review (Müller-Bloch and Kranz, 2015; Templier and Paré, 2018). We expect summarization, language translation, and question-answering capabilities of GenAI to provide useful support in each of these activities. These capabilities can enable teams to develop and assess different options for review projects. Going beyond the informal chartering activities, GenAI can compile evidence from the literature and offer an initial indication of which type of review aligns well with the current state of research. Risks of replicating existing reviews, possibly due to the use of non-standardized terminology or even due to publication in a different language, can be reduced by dedicated prompting strategies. In addition, once the review objectives are determined, GenAI may be applied to articulate the rationale for the review, position it relative to other review papers, and assemble the conceptual foundations and key definitions.

The example prompts illustrate how GenAI can be used to support the identification of prior review papers (Table 1), conceptual definitions (Table 2), and suitable review types (Table 3). Reasonable responses can be achieved when an initial set of relevant papers is provided as input, taking advantage of GenAI's PDF reading capabilities. In addition, it is advisable to add instructions specifying the expected output when nothing is found and to provide a clear definition of the review types in question. The latter could be done with reference to submission requirements at target journals (Rivard et al., 2018) and the methods discourse (Paré et al., 2015). We note that none of the example prompts can be answered based on titles and abstracts alone; all require an analysis of full-text papers.

### *Literature search*

The literature search commonly proceeds from an exploratory to a systematic search phase (Gusenbauer and Haddaway, 2021), and, in the context of GenAI, may increasingly intertwine with exploratory skimming and reading activities (Boell and Cecez-Kecmanovic, 2014; Palani et al., 2023; Wagner et al., 2020). One of the key challenges in the traditional process is that the massive volumes of research output resulting from literature searches quickly exceed human information processing capacities (Larsen et al., 2019). When researchers have limited prior knowledge of the literature, it is even harder to identify relevant papers and to direct exploratory reading activities. Emergent GenAI capabilities, like summarizing, classifying, and question-answering, may effectively enable researchers to overcome these limits, engage with the contents of larger sets of papers, and gain insights to adjust search activities. As such, we expect that GenAI capabilities can facilitate more pronounced exploratory search activities (Gusenbauer and Haddaway, 2021), complement strictly matching search strategies with semantic searches that include synonyms, as well as support the convergence between search and initial screening, skimming, and reading activities.

As an example of exploratory search capabilities, GenAI can be used for question answering or for generating summaries in the form of tables or graphs, providing researchers with a high-level overview of paper contents (Alshami et al., 2023). Table 4 illustrates this type of prompt. Promising online services in this area are often based on publicly available metadata, such as abstracts and open-access PDFs (e.g., Paperdigest[12] or scite.ai[13]) or even full-text documents of individual publishers (e.g., Scopus AI[14]). Additionally, researchers may take advantage of tabular summaries offered by services like Consensus,[15] or Elicit.[16] Providing access to, and requiring explicit connection to research papers enables these services to address the problems of hallucinations or fictitious references more effectively compared to early tests with generic GenAI tools (McGowan et al., 2023).

For systematic searches, which typically involve the design of Boolean search queries for academic databases, use of GenAI has major caveats but also promises to alleviate key challenges. Early experience reports repeatedly confirmed problems with hallucinations (McGowan et al., 2023), as well as responses that focus on openly accessible papers published in emergent outlets while missing most of the major contributions in the field. In addition, GenAI tends to lack access to paywalled content, recent publications, and unpublished work[17] potentially containing valuable findings on non-significant relationships. As such, few expect GenAI, such as ChatGPT, to replace the established retrieval process from academic databases in the near future.

Despite the shortcomings, GenAI offers a promising tool to facilitate and improve the design of systematic search strategies for established search infrastructure. Initial work indicates that GenAI performs well in handling (statistical) synonymy associations (Min et al., 2023; Thießen et al., 2023), which is one of the persistent challenges in finding prior research in many social science disciplines (Larsen and Bong 2016). In fact, identifying and grouping synonyms is at the core of constructing Boolean search queries following the building block approach, which refers to “dividing a query into Facets A, B, and C, complete with variants and synonyms, and then adding these concepts together using the Boolean AND operator” (Booth, 2008). In addition, general-purpose GenAI and specialized tools (e.g., DeepL) continue to improve in language translation tasks, offering the possibility to add terminology in different languages and with spelling variations to each concept block. Accordingly, initial research has evaluated the effectiveness of ChatGPT for writing Boolean search queries and found that results are particularly useful for reviews in which highly precise searches are acceptable, such as rapid reviews (Wang et al., 2023). As such, initial queries, such as those returned by the prompt example (Table 5), can already be used as a starting point, with further expansion of search terms needed to achieve adequate recall.

**Table 1.** Prompt to identify prior review papers based on citation context.

| | |
|---|---|
| GenAI-capability | Data extraction |
| Prompting strategy | Exploratory prompting |
| Requirements | LLMs with file upload and large context window (>100,000 tokens[a]) |
| Prompt example | **Upload a selection of relevant papers (PDFs)**[10] |
| | Considering the in-text citations of each paper, do the papers refer to prior (standalone) literature reviews? Which ones? Consider all PDFs and state explicitly when you encounter problems in extracting text from the PDF document. |
| Initial evaluation[11] | |
| Outcome | Successful identification of 5 out of 8 literature review papers (GPT-4o), 6 out of 8 literature review papers (Claude 3.5 Sonnet), and 5 out of 8 literature review papers (Gemini 1.5 Pro) |

[a]Tokens are the basic units of text that LLMs process; tokens can be short words (e.g., “the”) or parts of a word (e.g., “play” and “ing” in *playing*).

**Table 2.** Prompt to extract concept definitions.

| | |
|---|---|
| GenAI-capability | Data extraction |
| Prompting strategy | Zero-shot prompting |
| Requirements | LLMs with file upload and large context window (>100,000 tokens) |
| *Prompt example* | ***Upload a selection of relevant papers (PDFs)***[7] From the PDFs provided, extract the definitions for [add description here]. Provide a direct quote if remote work is defined in the paper. If there is none, state that there is no clear definition of [add topic label here] in the PDF. |
| Initial evaluation[8] | |
| Selected example | Remote work |
| Outcome | Successful extraction of 3 out of 5 definitions (GPT-4o), 3 out of 5 definitions (Claude 3.5 Sonnet), and 4 out of 5 definitions (Gemini 1.5 Pro) |

## *Literature screening*

In the literature screening phase, researchers label papers as relevant or irrelevant to the review, based on metadata or based on full-text documents (Templier and Paré, 2018). Given that only limited information (such as titles and abstracts) is available in the first screen, it is a good practice to retain borderline cases for the second screen. In the second stage, final inclusion decisions are made by examining the paper, and by documenting reasons for inclusion or exclusion in the form of screening criteria and reporting descriptive statistics on the screening process, for example, in line with the PRISMA standard (Page et al., 2021). The prevalent approach to controlling the reliability of the screening process is to have two (or more) researchers screen a sample redundantly, and to measure inter-coder reliability.

Initial research has explored the possibility of using GenAI for screening research papers, concluding that classification performance is currently not accurate enough to automate the process (Castillo-Segura et al., 2023; Syriani et al., 2024). Specifically, the work of Syriani et al. (2024) reports how the performance of GPT-3.5, using the best performing prompt template displayed in Table 6, compared to the traditional AI classifiers. While the consistency of screening decisions for individual records was relatively high, it is noteworthy that metrics for LLM-based classification vary considerably across datasets, indicating limited generalizability. In particular, the findings show that the essential recall metric does not dominate the performance of random classification in all cases. As such, follow-up research is needed to determine whether larger models or different prompting strategies can improve classification performance. More generally, it is important to remember that language-based AI techniques are not the only ways to automate and facilitate literature reviews. Traditional AI technologies, such as those based on in-house trained neural networks (Wagner et al., 2022) may be a better choice, especially when high accuracy and reliability are needed.

**Table 3.** Prompt to assess the fit of a selected review type.

| | |
|---|---|
| GenAI-capability | Text analysis and recommendations |
| Prompting strategy | Exploratory prompting |
| Requirements | LLMs with file upload and large context window (>100,000 tokens) |
| Prompt example | **Upload a selection of relevant papers (PDFs)**[7] Definition: A “qualitative systematic review” aims at collecting and aggregating empirical evidence from primary studies to test a narrowly defined hypothesis or model. It is suitable for established topics for which the research questions are narrow and the focus is on qualitative or quantitative empirical studies. Assess a review project focusing on [add description here]. Would a qualitative systematic review be a suitable review type? Provide reasons related to the topic maturity, the scope of research questions, and the nature of prior work. |
| Initial evaluation[8] | |
| Selected example | Future of work |
| Outcome | Convincing identification and justification of 2 out of 5 review types (GPT-4o), 1 out of 5 review types (Claude 3.5 Sonnet), and 1 out of 5 review types (Gemini 1.5 Pro) |

**Table 4.** Prompt to explore prior research using a tabular overview.

| | |
|---|---|
| GenAI-capability | Text summarization |
| Prompting strategy | Retrieval-augmented generation (RAG) |
| Requirements | LLMs with Retrieval Augmented Generation functionality, such as Consensus or Elicit |
| Prompt example | How does [add label of variable] affect the relationship between [add antecedent variable or intervention] and [add outcome variable] in the context of [add context description]? Summarize relevant empirical papers with an abstract summary, the research method, and the key findings |
| Initial evaluation[8] | |
| Selected example | Effect of skills on the relationship between LLM support and individual productivity in the context of software development |
| Outcome | From the first 10 papers returned, 2 were relevant and 8 not relevant (Elicit), 2 were relevant, and 8 not relevant (Consensus). All summaries were adequate and almost exclusively based on the abstracts. Concerning the nature of sources, Elicit returned 4 preprints and 2 papers from reputable journals. Consensus returned 2 preprints and 4 papers from reputable journals |

Against this background, we expect that GenAI may find a range of applications in support of the screening process. First, in the case of rapid reviews (common in non-scientific outlets), it may be acceptable to trade-off recall against quick completion of the review process, for instance to inform quick and informal (Syriani et al., 2024) and non-mission-critical (Lukyanenko et al., 2025). Second, the capabilities of GenAI may be particularly suitable to complete large-scale reviews. As such, we may see examples complementing the work of Larsen et al. (2019) to cover review topics that do not focus on a particular theoretical model with distinct constructs. Third, GenAI can facilitate a range of preparation tasks, including the development of training materials, examples, and process documentation for coders. These can be later shared to increase transparency and replicability (Burton-Jones et al., 2021; Hevner et al., 2024). Fourth, capabilities of translation can be particularly useful in the screening activities to overcome prevalent language and geographical biases (Van Wee and Banister, 2023), as suggested in Table 7. Fifth, GenAI can be applied to implement publication filters, such as restrictions to empirical studies required in meta-analyses. Given that researchers often use catchy, rather than descriptive titles,[18] it may be practical to apply such filters in the full-text screening stages rather than in the search (Higgins et al., 2023). Sixth, GenAI-based screening results can be used for parallel independent reliability assessment, especially in single-authored review papers (Templier and Paré, 2018), or for prioritizing screening activities (Syriani et al., 2024; Van de Schoot et al., 2021). Finally, text summarization may even allow researchers to modify screening criteria and understand whether and how conclusions would change. If effective strategies of using GenAI for this purpose can be developed, this would allow researchers to complete qualitative robustness checks and validate some of the more challenging and consequential methodological choices. In addition, such work could show how screening criteria emerge from a mutually informative process iterating between humans and GenAI-based machines, as well as search, screen, and reading activities (Boell and Cecez-Kecmanovic, 2014).

**Table 5.** Prompt to suggest an initial search query.

| | |
|---|---|
| GenAI-capability | Content generation |
| Prompting strategy | Few-shot prompting |
| Requirements | LLMs |
| Prompt example | You are an information specialist who develops Boolean queries for systematic reviews. You have extensive experience developing highly effective queries for searching the information systems literature. Your specialty is developing queries that retrieve as few irrelevant documents as possible and retrieve all relevant documents for your information needs. You are able to take an information need such as: "Review of IT Business Value" and generate valid Web of Science queries such as:<br>"TI= (IT OR IS OR …) AND TI= (value OR payoff OR …) AND TI= (firm OR business OR …)".<br>Now you have your information need to conduct research on "[add topic here]", please generate a highly effective systematic review Boolean query for the information need. |
| Initial evaluation[8] | |
| Selected example | Effect of LLM on individual performance at work |
| Outcome | Best performing prompt (without examples) out of five prompts as evaluated on GPT-3.5 (Wang et al., 2023) |

**Table 6.** Prompt to screen papers based on title and abstract.

| | |
|---|---|
| GenAI-capability | Text analysis and recommendations |
| Prompting strategy | Zero-shot prompting |
| Requirements | LLMs |
| Prompt example | **Extract abstracts locally and provide them with the prompt**[7]<br>Context: I am screening papers for a systematic literature review. The topic of the systematic review is [add topic here]. The study should focus exclusively on this topic<br>Instruction: Decide if the article should be included or excluded from the systematic review. I give the title and abstract of the article as input. Only answer include or exclude. Be lenient. I prefer including papers by mistake rather than excluding them by mistake<br>Task i:<br>- Title: "Twelve tips to leverage AI for efficient and effective medical question generation"<br>- Abstract: "Crafting quality assessment questions in medical education […]" |
| Initial evaluation[8] | |
| Selected example | Effect of generative AI on individual productivity for programmers |
| Outcome | Best performing prompt that maximizes F2 scores as evaluated on GPT-3.5 (Syriani et al., 2024) |

### Quality assessment

Formal quality assessment is particularly relevant for theory-testing reviews, including meta-analyses and qualitative systematic reviews (Templier and Paré, 2018). GenAI may assist with basic evaluations related to the methodological aspects of research studies, including the analysis of study designs, sample sizes, data collection methods, and statistical techniques (see Table 8, for an example). Future work may also show whether these models can be used effectively to identify potential flaws, biases, or limitations in the selected studies and flag them for further review by human researchers. Additionally, GenAI can help identify questionable research practices or logical errors (Habernal et al., 2018). It may also be leveraged to augment manual human quality assessments by ensuring consistency across multiple reviewers. By analyzing the assessments provided by different reviewers, the AI model can identify discrepancies or inconsistencies in the evaluation criteria or ratings, allowing for resolution and alignment.

In the future, one potential application of GenAI may be conducting parallel independent assessments of study quality. The methodological literature recommends multiple independent assessors evaluate the quality of studies included in a review (Templier and Paré, 2018). GenAI models, with few-shot prompting, can perform these independent assessments, providing an additional layer of evaluation alongside human reviewers (Weber, 2024). Furthermore, GenAI models could play a role in identifying, and refining the criteria used for quality assessment. By analyzing large datasets of literature reviews and their associated quality assessments, this involves identifying patterns or best practices in the assessment criteria and processes.

**Table 7.** Prompt for language translation in the screening process.

| | |
|---|---|
| GenAI-capability | Text translation |
| Prompting strategy | Zero-shot prompting |
| Requirements | Use GROBID to convert PDF documents to TEI format and provide the TEI (XML) files as an input to the LLM[7,19] |
| Prompt example | Read each XML document, which has the namespace https://www.tei-c.org/ns/1.0<br>Extract the following items:<br>- Title, which is in TEI/teiHeader/fileDesc/titleStmt/title (display in title case)<br>- Abstract, which is in TEI/teiHeader/profileDesc/abstract/div (using all p tags)<br>- Keywords, which are in TEI/teiHeader/profileDesc/textClass/keywords<br>Translate the abstract to English (if necessary)<br>Arrange all results in a Markdown table. Add a "screening" column. |
| Initial evaluation[8] | |
| Outcome | Successful translation of 5 out of 5 abstracts (GPT-4o), 5 out of 5 abstracts (Claude 3.5 Sonnet), and 5 out of 5 abstracts (Gemini 1.5 Pro) |

**Table 8.** Prompt for the basic evaluation of the methodological approach.

| | |
|---|---|
| GenAI-capability | Text analysis and recommendations |
| Prompting strategy | Zero-shot prompting |
| Requirements | LLMs with file upload and large context window (>100,000 tokens) |
| Prompt example | **Upload a paper (PDF)**[7]<br>Your task is to analyze the provided research study and identify its study design (e.g., experiment, case study, survey, archival study), sample size, data collection methods, and statistical analyses. Present these four characteristics in a Markdown table |
| Initial evaluation[8] | |
| Outcome | Convincing evaluation of 2 out of 5 methodological approaches (GPT-4o), 1 out of 5 methodological approaches (Claude 3.5 Sonnet), and 1 out of 5 methodological approaches (Gemini 1.5 Pro) |

### Data extraction

In the data extraction activities, it is particularly instructive to consider the *jagged frontier* of GenAI. This refers to the observation that, instead of leading to consistent improvements across tasks, GenAI can have unpredictable effects when "tasks that appear to be of similar difficulty may either be performed better or worse by humans using AI" (Dell'Acqua et al., 2023: 8). For example, while GenAI may perform reliably when extracting explicit characteristics, such as sample sizes or participant demographics, yet fail in closely related situations requiring greater interpretive judgment, such as inferring an author's implicit epistemological stance or interpreting complex robustness checks in the reported results.

For descriptive reviews, where the goal is to summarize research findings, GenAI has already demonstrated its effectiveness in creating concise and accurate summaries. Importantly, these AI-generated summaries can be tailored to the researcher's specific needs, focusing on particular themes or methodological characteristics, or adjusted to suit different target audiences (see Table 9, for an example).

While GenAI may be less useful for data extraction in review aimed at theory development, it shows great promise for theory-testing reviews, such as meta-analyses, which rely on structured data extraction. GenAI can extract structured data from text, including study characteristics (Dagdelen et al., 2024), and potentially correlations and effect sizes from primary studies. These GenAI classification capabilities may be most useful for small or emergent subject areas where no validated ML models for classification exist. Before GenAI models, developing a text classifier for a non-standard problem, including the annotation of a training dataset, required a substantial amount of work. With GenAI, getting up and running with a classifier may become much easier (see Table 10, for an example).

As GenAI continues to improve, with expanding context windows and enhanced ability to generate structured and reproducible output (Dagdelen et al., 2024), we can envision GenAI automating even more complex data extraction tasks. For instance, it may become possible for GenAI to extract correlation tables, compile effect sizes from a sample of primary studies, and assist in automatically conducting meta-analyses. This could significantly reduce the time and effort required for theory-testing reviews (Li et al., 2026), allowing researchers to focus on other steps of the process or even entirely different types of reviews with more substantial interpretation and synthesis requirements.

### Data analysis

In data analysis and code development tasks, GenAI has demonstrated remarkable capabilities (Peng et al., 2023) and tools like MAXQDA have started to integrate LLM capabilities. Generally, one of the most popular use cases of GenAI—text generation—may be leveraged across all types of reviews, for example, to draft descriptive summaries based on research papers in narrative, descriptive or scoping reviews as well as to assist in editing and proofreading tasks for theoretical reviews (Huang and Tan, 2023; Skarlinski et al., 2024). Although some have argued that purely text-generative tools may not be capable of supporting evidence-aggregating studies (Rahman et al., 2023; Schryen et al., 2024), literature reviews that involve structured analyses such as meta-analyses may benefit from GenAI by developing code for analyses or data visualizations (see Table 11, for an example). Perhaps most significantly, recent work leverages GenAI to support dynamic, real-time theory testing reviews that automatically update as new studies are published, ensuring researchers always have a synthesis of the latest evidence available (Li et al., 2026). In this manner, GenAI can unleash a new form of publishing whereby a paper is a living document, updated in-vivo (on the publishing platform) as new evidence emerges.

On the other end of the spectrum, for theory-building reviews with less strict data analysis schemas and inductive analyses, the support from AI may be limited to writing assistance, as these tasks require a higher level of conceptual understanding and idea generation. Nonetheless, the interactional quality of chatbot implementations of GenAI can support idea generation and refinement through, for example, Socratic-style argumentation about emerging theoretical ideas which may satisfy criteria for "the epistemic community values of argumentation" (Ngwenyama and Rowe, 2024: 123), as illustrated in the example prompt below (see Table 12). Looking ahead, GenAI will play a more substantive role especially in inductive, theory-building work by acting as a co-researcher that summarizes the researcher's memos identifying gaps in understanding, identifying key concepts in research papers and mapping conceptual relationships. It could thereby complement existing approaches primarily applicable to hypothetico-deductive research (Li et al., 2020).

**Table 9.** Prompt for chain-of-density summarization.

| | |
|---|---|
| GenAI-capability | Text summarization |
| Prompting strategy | Chain-of-thought prompting |
| Requirements | LLMs with file upload and large context window (>100,000 tokens) |
| Prompt example | **Upload a paper (PDF)**[7]<br>You will generate increasingly concise, entity-dense summaries of the above article. The summaries should be written for an academic audience.<br>Repeat the following 2 steps 5 times<br>- Step 1. Identify 1–3 informative entities (";" delimited) from the article which are missing from the previously generated summary.<br>- Step 2. Write a new, denser summary of identical length which covers every entity and detail from the previous summary plus the missing entities.<br>A missing entity is:<br>- Relevant: to the main story<br>- Specific: descriptive yet concise (5 words or fewer)<br>- Novel: not in the previous summary<br>- Faithful: present in the article<br>Anywhere: located anywhere in the article |
| Initial evaluation[8] | |
| Outcome | Best performing prompt (after four chain of density steps) that maximize entity density and surpass human summaries (Adams et al., 2023) |

**Table 10.** *Python* pseudocode for structured data extraction from tables.

| | |
|---|---|
| GenAI-capability | Data extraction |
| Prompting strategy | Zero-shot prompting |
| Requirements | LLMs with file upload and large context window (>100,000 tokens) |
| Prompt example | **Upload a paper (PDF)**[7]<br>1. Define utility functions:<br>• md_to_df(markdown_text): Converts Markdown table text to a pandas DataFrame<br>• extract_table_from_image(url): Extracts table data from an image at the given URL and returns as Markdown text<br>2. Define the MarkdownDataFrame data structure:<br>• Use pandas.DataFrame as the base structure<br>• Apply a BeforeValidator that converts Markdown text to a DataFrame (md_to_df function)<br>• Apply a PlainSerializer to convert a DataFrame to Markdown text (using DataFrame.to_markdown() method)<br>• Define JSON schema for validation<br>3. Define the Table class with two attributes: caption and dataframe:<br>• Caption: String to store the table's caption<br>• dataframe: Stores the table data as a MarkdownDataFrame, which is essentially a pandas DataFrame that can serialize to/from Markdown.<br>4. Main process to extract and represent a table from an image:<br>• Call extract_table_from_image(url) to extract the Markdown representation of the table from the image<br>• Create an instance of the Table class, setting caption as needed and dataframe as the Markdown representation converted to a DataFrame<br>• Use the Table instance to manipulate or access the table's data and caption<br>• To serialize the Table instance's dataframe back to Markdown, use the PlainSerializer functionality implicitly via the class's structure |
| Initial evaluation[8] | |
| Outcome | Successful data extraction from 4 out of 5 tables (GPT-4o), 5 out of 5 tables (Claude 3.5 Sonnet), and 4 out of 5 tables (Gemini 1.5 Pro) |

**Table 11.** Prompt to develop *Python* code for a meta-analysis.

| | |
|---|---|
| GenAI-capability | Code generation |
| Prompting strategy | Zero-shot prompting |
| Requirements | LLMs |
| Prompt example | As a *Python* programming and statistical analysis expert with a detailed understanding of conducting meta-analysis in *Python*, you are tasked with generating *Python* code that aligns with the following steps:<br>- Step 1: Install the PythonMeta (V.1.26) package and read a dataset. The dataset is sitting in the same file directory as the *Python* scripts<br>- Step 2: Generate main results by selecting binary outcome and Risk Ratio as the desired effect size. Run both fixed-effect and random-effects models, choosing MH for fixed-effect and DL for the random-effects models. Generate forest plots and funnel plots<br>- Step 3: Assess the impact of missing data. After cleaning the dataset, label the studies with missing and non-missing patients and analyze them as subgroups. Implement missing data imputation methods including Available Case Study (ACS), Imputed Case Analysis (ICA), and best and worst-case scenarios. Run a separate random-effects model with IV method on each and generate relevant forest plots<br>- Step 4: Evaluate the small study effect, assess the asymmetry of the funnel plots, and perform Egger's test using Statsmodels linear regression<br>Remember to format the responses in a clear and precise format. Output tables when possible. Keep your tone professional and instructional, ensuring the generated Python code adheres to best practices for readability and efficiency |
| Initial evaluation[8] | |
| Outcome | Generation of working meta-analysis code with minor fixes (GPT-4o), working meta-analysis code on first attempt (Claude 3.5 Sonnet), and working meta-analysis code with minor fixes (Gemini 1.5 Pro) |

## Reflections, opportunities, challenges, and open questions

The capabilities of GenAI to assist with literature reviews are already impressive and continue to improve, as companies and even countries begin to compete to create better foundational and specialized GenAI models. At the same time, important questions and opportunities related to methodological and technological challenges and the future of scientific progress must be raised, to ensure the use of GenAI tools for literature reviews is effective, but also responsible. In this section we aim to present a balanced outlook by highlighting the positive potential of GenAI for literature reviews while also critically questioning some developments that could undermine long-term scientific innovation and creativity or pose risks from opening up scientific processes too broadly. In the following, we summarize our own findings, thereby providing a backdrop for more long-term reflections, including the potential impact of GenAI on scientific progress, types of review and the technological challenges of GenAI that should be overcome to unlock even greater potential of this transformative technology.

### *Discussion of our findings*

Despite the caveats and limitations, our analysis reveals considerable promises of GenAI, which is highly capable of augmenting and even, in some cases, entirely automating activities of the literature review process. Both generic (e.g., ChatGPT, Claude, Gemini) and specialized (e.g., Consensus, Elicit) tools should be actively considered by researchers on most review projects. At the same time, the tools are best thought of as methodological co-pilots, rather than wholesale replacements of manual human effort.

**Table 12.** Prompt to reframe theoretical questions based on Socratic argumentation.

| | |
|---|---|
| GenAI-capability | Dialogue and conversation |
| Prompting strategy | Exploratory prompting |
| Requirements | LLMs with file upload and large context window (>100,000 tokens) |
| Prompt example | **Upload a selection of relevant papers (PDFs)**[7]<br>You are an AI assistant capable of having in-depth Socratic style conversations on a wide range of topics. Your goal is to ask probing questions to help the user critically examine their beliefs and perspectives on the attached paper. Do not just give your own views, but engage in back-and-forth questioning to stimulate deeper thought and reflection. |
| Initial evaluation[8] | |
| Outcome | Adapted from best performing prompt (without extra knowledge) that involves structured conversation, encompassing review, heuristic, rectification, and summarization (Ding et al., 2024) |

Some activities of the literature review process appear to be especially amenable to support or in some cases, full automation, with GenAI. Thus, problem formulation can be greatly enhanced by GenAI as it excels at navigating ambiguity, providing a sense of present-day developments and improving conceptual understanding of early-stage literature exploration, at an unprecedented scale. Similarly, for literature searches, GenAI is particularly apt at supporting or supplementing exploratory activities.

For some activities, GenAI can be incredibly helpful, but often requires careful and precise prompts, and does not always outperform manual or traditional AI-driven initiatives. If the aim is to maintain maximal transparency and controls, manual effort or carefully designed and meticulously validated (Ethayarajh and Jurafsky, 2020; Larsen et al., 2025) custom AI classification models should be preferred (Wagner et al., 2022). Here, we observe a perennial tension between rigor and scale, known in other settings, as for example, the trade-off between accuracy and completeness, precision and recall, or internal and external validity.

The generic abilities of GenAI to summarize content at scale, and transform content presentation (e.g., creating tables, graphs), enhance researchers' ability to understand and communicate the findings. The tools also permit the evolution and enhancement of the methods underlying literature reviews. For example, GenAI may permit qualitative robustness checks and hence validate some of the methodological choices, which are rarely validated at the moment (e.g., the search strategy or screening criteria). Similarly, GenAI may be used to conduct parallel independent data extraction, quality control, screening and search activities which can be compared to manual efforts. Incorporating these possibilities into literature review methods is an exciting frontier for research.

The present-day capabilities and even greater future potential of GenAI have profound implications for how GenAI may shape the trajectory of scientific progress in both beneficial and detrimental ways that need to be carefully managed. Next, we consider the broader effect on long-term scientific innovation and progress.

## GenAI and literature review types

Although GenAI shows considerable potential, its utility varies significantly depending on the type of literature review and the specific demands of each review stage. To explore this, we present four scenarios—*Obsolescence*, *Varying Degrees of Augmentation*, *Inadequacy*, and *New Trajectories*—each representing a unique pathway through which GenAI could transform literature review practices. These scenarios highlight GenAI's impact on different types of reviews, the changes it may introduce to specific review steps, and the broader implications for literature review methodologies.

In the *Obsolescence* scenario, GenAI advancements lead to the partial or complete replacement of certain types of literature reviews. Specifically, GenAI's summarization, automated synthesis, and bibliometric capabilities make some reviews—such as descriptive reviews and bibliometric studies—obsolete, particularly those focused on categorizing and summarizing existing literature. GenAI's ability to rapidly collect, classify, and synthesize large datasets reduces the need for manual effort in these review types. The activities most affected by this scenario include search and screening, where GenAI may fully automate or significantly streamline these processes, and synthesis, where AI can categorize and summarize literature with minimal human intervention.

The *Varying Degrees of Augmentation* scenario captures the spectrum of GenAI's potential impact on literature reviews, where GenAI serves as a supportive tool rather than a replacement. In this scenario, GenAI capabilities are selectively applied based on the complexity and demands of each review stage. For instance, narrative reviews, scoping reviews, and meta-analyses may benefit from GenAI in tasks like exploratory searches, screening, and data extraction, while human oversight remains essential for data synthesis and interpretation to maintain rigor and accuracy. GenAI's role in this scenario is to enhance traditional review processes by increasing efficiency and reducing researchers' time demands, allowing them to focus on high-level analysis and interpretation. This collaborative model underscores the need for researchers to retain methodological expertise while leveraging AI effectively.

In the *Inadequacy* scenario, GenAI tools, despite their capabilities, remain insufficient for certain types of literature reviews. Reviews requiring substantial theoretical, conceptual, or interpretive synthesis—such as theoretical reviews, conceptual reviews, meta-narrative reviews, meta-ethnographies, and critical reviews—demand deep contextual understanding and interpretative skills that GenAI currently lacks. Here, GenAI might minimally assist in exploratory search and preliminary reading, but it falls short in synthesis and critical interpretation. These review types depend on researchers' interpretative lenses and subjective insights, which cannot be replicated by generative models alone. This scenario underscores GenAI's limitations, particularly when interpretative depth and nuanced analysis are required, and reinforces the ongoing importance of human expertise in qualitative and theoretical reviews. It highlights the value of "human-in-the-loop" models, where AI aids in preliminary tasks but requires significant human oversight for rigorous interpretation.

Finally, the *New Trajectories* scenario envisions GenAI as a catalyst for innovation, enabling novel types of literature reviews or expanding the scope of traditional reviews in ways previously infeasible. This scenario is especially relevant for interdisciplinary reviews that are simultaneously broad and in-depth, and those where researchers apply established theories to new, unrelated contexts (Gregor and Hevner, 2013). GenAI's potential for translational capabilities—its ability to bridge disciplinary knowledge and generate cross-disciplinary insights—allows researchers to engage with literature beyond their immediate field, fostering a new level of interdisciplinary integration. Moreover, GenAI could enable radically scaled review efforts, allowing researchers to analyze vast amounts of literature from multiple disciplines efficiently. By facilitating knowledge transfer across fields, this scenario could contribute to theoretical advancement and the development of interdisciplinary frameworks. However, it also requires researchers to remain vigilant in ensuring the accuracy and relevance of insights, particularly when working in unfamiliar domains.

In summary, these four scenarios illustrate the diverse ways in which GenAI could impact literature reviews. From the automation of descriptive reviews to selective augmentation in theory-testing reviews, limited applicability in interpretative reviews, and new opportunities in interdisciplinary synthesis, each scenario underscores a different facet of GenAI's transformative potential. As researchers navigate these possibilities, they must balance GenAI's efficiencies with critical oversight, ensuring methodological rigor and adapting to the evolving landscape of AI-supported research.

Drawing on established classifications of review types (Paré et al., 2015, 2024; Rowe, 2014; Schryen et al., 2020), Table 13 provides an at-a-glance overview of how each review type might leverage GenAI, noting that some types align better with specific scenarios than others. For instance, GenAI can significantly enhance meta-analyses by supporting systematic tasks such as data extraction, writing code for the meta-analytic regressions, and preliminary synthesis, aligning well with the *Varying Degrees of Augmentation* scenario. GenAI's ability to automate data handling processes, identify relevant studies, and summarize results increases the efficiency of meta-analyses, allowing researchers to focus on higher-level analysis and interpretation. However, for more complex synthesis and interpretation, particularly when assessing study heterogeneity or addressing nuanced methodological issues, human expertise remains essential. Therefore, while GenAI can streamline many aspects of meta-analyses, rigorous oversight and critical evaluation by researchers are still required to ensure accuracy and robustness in the final synthesis. As another illustration, GenAI can augment certain stages of critical reviews by helping with systematic aspects, such as locating and summarizing literature. However, the interpretive depth, evaluative focus, and critical perspective that define critical reviews place them primarily within the *Inadequacy* scenario, as these tasks require sophisticated human judgment (Block and Kuckertz, 2024). Thus, while GenAI may support some activities, the core critique remains a human-centered task.

### Addressing technological challenges of GenAI

While GenAI continues to impress with its capabilities, it also sometimes disappoints. In order to pave the way for even greater impact, several key limitations of modern GenAI need to be overcome. Considering our analysis of GenAI for literature reviews, we suggest fruitful opportunities for research that seeks to improve GenAI itself. We focus on two central issues: architectural and data-related challenges as areas of future research.

### Architectural challenges

Limitations of GenAI due to architectural issues of this technology pose significant hurdles for their effective application for literature reviews. One of the foremost challenges is the propensity of these models to produce hallucinations, that is, generating information that is factually incorrect or not present in the source data. In the context of literature reviews, such hallucinations can lead to misrepresentation of research findings, citation of non-existent studies, or incorrect summarization of key concepts, thereby compromising the integrity of the review. To mitigate these issues, techniques like RAG have shown promising results (Li et al., 2024). RAG enhances factual accuracy by enabling models to access and reference external databases during generation. Another promising approach is the curation of knowledge graphs representing literature sources. In this approach, some of the especially critical semantics does not have to be extracted from literature sources and can be embedded directly. For example, statistical information reported (e.g., coefficients of structural equation models), or specific definitions (e.g., constructs, relationship among constructs), can be directly represented as knowledge graphs, ensuring precise incorporation of this information into GenAI representations. To support these developments, research is needed across a wide spectrum, ranging from the efficient collection and management of knowledge graphs (e.g., potentially supported by community-curated repositories, crowdsourcing, and other human in the loop approaches to ensure high accuracy of ground-based representations), along with continued work on graph-based knowledge embedding in large language models (Pan et al., 2024).

Understanding nuanced contexts and critical synthesis prevalent in academic literature are significant technological challenges that GenAI models struggle with. Conducting literature reviews not only requires a deep comprehension of domain-specific language (which can be provided with appropriate data and methods like RAG—as we noted before) and theoretical frameworks, but also requires critical thinking and reasoning. Studies in this domain have shown that GenAI models, while capable of analogical and moral reasoning, struggle with other reasoning tasks such as spatial reasoning (Agrawal, 2023). General-purpose AI models might not capture important subtleties, leading to superficial synthesis or misinterpretations of critical concepts. This limitation hinders the ability of AI to fully assist in synthesizing complex scholarly work and may necessitate significant human oversight to correct and refine the outputs. As a remedy, specialized models trained on domain-specific corpora should be developed to address the issue of nuanced understanding. By tailoring models to specific fields, researchers can improve the models' grasp of specialized terminology and complex concepts. Additionally, transfer learning and other approaches such as reinforcement learning with human feedback (RLHF) enable models to improve their abilities in critical thinking and reasoning. These approaches have resulted in more recently developed GenAI models such as OpenAI's O1 Preview, which is shown to substantially outperform humans in "systematic thinking, computational thinking, data literacy, creative thinking, scientific reasoning, and abstract reasoning." (Latif et al., 2024). Furthermore, recent research has found that the performance of the O1 model, which was developed utilizing advanced reinforcement learning techniques that significantly surpass traditional RLHF methods, consistently improves with increased reinforcement learning during training (train-time compute) and with more time allocated for reasoning during inference (test-time compute) (Latif et al., 2024).

**Table 13.** Alignment of main review types with GenAI integration scenarios.

| Overarching goal | Review type | GenAI integration scenarios | | | |
|---|---|---|---|---|---|
| | | Obsolescence | Augmentation | Inadequacy | New trajectories |
| Describing | Narrative | Unlikely | Moderate—supports search and summarization | Likely—requires interpretive synthesis | Unlikely |
| | Descriptive | Likely—potential for automation | Moderate—thematic extraction | Unlikely | Unlikely |
| | Scoping/mapping | Moderate—structured tasks | Likely—systematic search and categorization | Unlikely | Unlikely |
| Theory testing | Meta-analysis | Moderate—data extraction, synthesis | Likely—supports search, data extraction, and statistical synthesis | Moderate—requires oversight in complex synthesis | Moderate—potential for cross-disciplinary integration |
| | Systematic | Moderate—structured tasks | Likely—supports search, screening, and synthesis | Unlikely | Unlikely |
| | Umbrella | Moderate—high-level synthesis | Likely—search, screening, and summarization | Unlikely | Moderate—interdisciplinary synthesis |
| | Rapid | Moderate—high-level synthesis | Likely—prioritizes speed, search, screening | Unlikely | Unlikely |
| Theory building | Theoretical | Unlikely | Moderate—supports thematic organization | Likely—requires theoretical synthesis and interpretation | Moderate—potential for concept translation across disciplines |
| | Realist | Unlikely | Moderate—supports search, data extraction | Likely—requires context-sensitive interpretation | Moderate—future potential in contextual integration |
| Understanding | Meta-narrative | Unlikely | Likely—supports thematic organization | Likely—requires interpretative synthesis across narratives | Moderate—interdisciplinary knowledge translation |
| | Critical | Unlikely | Moderate—supports initial stages | Likely—requires critical interpretive analysis | Unlikely |
| | Problematization | Unlikely | Moderate—supports initial stages | Likely—requires questioning assumptions | Unlikely |

The lack of interpretability and transparency inherent in many AI models (such as deep learning–based LLMs) is a significant technological challenge. GenAI models often function as "black boxes," making it difficult for researchers to trace how specific outputs are generated from given inputs. This opaqueness is problematic in academic settings where the justification of conclusions and the reproducibility of results are essential. Researchers may find it challenging to trust the insights provided by AI models if they cannot understand the models' reasoning processes, which undermines the utility of these tools in conducting rigorous literature reviews. To address this issue, developments in explainable AI (XAI) are enhancing the interpretability of model outputs by offering insights into the decision-making processes of AI systems (Swamy et al., 2024). For instance, attention mechanisms and gradient-based attribution methods allow researchers to identify which parts of the input data the model focuses on when generating responses. This transparency helps researchers understand and trust the AI's contributions to literature reviews.

Handling multi-modal data introduces additional technological complexities. Multi-modal GenAI aims to process and integrate information from various sources such as text, images, graphs, and tables, which are commonly found in academic articles. However, effectively combining these different data modalities to generate coherent and meaningful analyses remains a significant challenge. The models may not accurately interpret visual data like charts or may fail to correlate information across modalities, resulting in incomplete or biased literature reviews. In the realm of multi-modal data processing, innovative architectures like Transformers with modality-specific encoders are improving the integration of diverse data types. Models such as OpenAI's CLIP (Contrastive Language-Image Pre-training) demonstrated promising performance in associating textual and visual information. These advancements can enhance the AI's ability to interpret and synthesize information from different formats commonly found in academic literature.

Computational resource demands also present a barrier. The sophisticated architectures of GenAI models require substantial processing power and memory. This requirement can limit accessibility for individual researchers or institutions with constrained resources, thereby impeding widespread adoption of these technologies in academia. The high costs associated with training and deploying such models can also divert funding from other critical research activities. Efforts to reduce computational requirements are also underway. Techniques like model pruning (Ma et al., 2023), quantization (Egashira et al., 2024), and knowledge distillation (Xu et al., 2024) help create smaller, more efficient models without significantly sacrificing performance. These approaches make it more feasible for researchers with limited resources to utilize advanced AI tools.

### *Data-related challenges*

As with any other data intensive artificial intelligent technology, the issues pertaining to data play an outsized role in the continued maturity of GenAI. There are many challenges and opportunities related to data management for GenAI in the context of literature reviews.

One of the significant issues is data access. A promise of GenAI for literature reviews is in its ability to expand the coverage of topics beyond what is humanly possible. However, the realization of this promise is being impeded by the inability of present tools to capture the entirety of the relevant literature. As a result, any literature review findings or analyses would be biased toward available sources. What is worse, some of the sources (e.g., ArXiv.org) while being relevant, may not guarantee a rigorous peer review process, and therefore may not be as reliable as the carefully curated sources in the inaccessible databases.

Design science researchers can address the many data-related issues from a multitude of perspectives. First, an opportunity exists to improve the GenAI's development routine to automatically ascertain the quality, bias, and representativeness of the sources used (Parsons et al., 2025; Zhang et al., 2019), permitting the AI models to better leverage the training data in generating the responses. Effectively, this is the concept behind retrieval augmented generation (RAG), except the research focus here is on the upstream part of the AI training, such that the tools would become more sensitive to the varying levels of data quality and representativeness. This issue, while a general one, is especially important for literature reviews, as relevant to a research question literature can drastically vary in its quality. Considering this variability, GenAI tools can offer different analyses, depending on the sensitivity of the research team to the sources and their quality levels (e.g., analysis on the entire available corpus, only the most reputable sources, gray literature, etc.).

Second, data management scholars can support the GenAI industry with solutions that can lessen the monetary burden of having to procure sources from paid databases. These approaches, for example, can draw upon research on differential privacy (Dwork, 2006) and information obfuscation (Liao et al., 2021), where only relevant information is made accessible and shared, to minimize such concerns related to copyright and intellectual property protection and reduce the information transfer volume.

Third, even scientific literature in highly curated, paid databases, is not necessarily bias free. Weber (2024), when considering AI as a tool for reviews, warns scientific disciplines exhibit often hard-to-detect entrenched biases. An important opportunity therefore is to develop systematic techniques to automatically identify these biases and make researchers aware of them. This work can leverage growing research on AI data bias identification and mitigation (Chen et al., 2024; Nazer et al., 2023; Tejani et al., 2024). Despite much progress, one overlooked opportunity is communicating bias to the user through a user interface, and ensuring that the user (e.g., scientist-in-training), knows how to appropriately account for the biases in the literature.

Finally, there is an important novel opportunity related to what we call *prompt data management*. Prompt data management is a new data management frontier that focuses on collection, curating, classification, and support for usage of effective prompts for GenAI. In our context, these prompts are tailored to literature reviews. Prompt curation requires its own considerations, different from the generic data management contexts. As we discussed and showed in our paper, in addition to curating the text of the prompt, certain properties and details of the prompt are important to capture and curate. Effective prompts for GenAI follow patterns which are not yet well-established and understood. For example, prompts for literature review are often required to be issued in a particular sequence (as literature search is a complex and multi-phased process). Hence a challenge of prompt data management is understanding the patterns of effective prompts for literature reviews, classifying them appropriately so users can easily find those needed for their tasks, and developing accessible repositories for such prompts. To begin realizing this vision, we created a repository of literature review prompts, which will be updated continuously with recent prompts that are published and evaluated in scientific outlets.[20] Future research can study prompts data management to better understand and refine the practices for collecting and curating literature review prompts.

In conclusion, although some technological challenges may currently limit the full potential of GenAI for literature reviews, ongoing advancements and innovations are steadily overcoming these obstacles. As the technology matures, we can anticipate more reliable, interpretable, and accessible GenAI models that will significantly enhance the efficiency and depth of literature review.

## General open questions

There are many open questions, beyond design of GenAI and the identification of effective prompts, including standards for human oversight, and reporting principles. More fundamentally, new answers are needed on how GenAI can enrich human understanding. Accordingly, researchers should explore possibilities of bringing GenAI to hermeneutic traditions and theory development reviews, fostering a new era of interdisciplinary research that bridges the gap between computational analysis and human interpretation. At the top of the human cognitive ability pyramid lie creativity, critical thinking, and complex problem solving. When paired with GenAI capabilities, these skills have the potential to enhance our understanding, interpretation, and synthesis of prior knowledge.

Careful attention should also be given to the misuse of GenAI (cf. Susarla et al., 2023). In the context of literature reviews, the recent work of Tingelhoff et al. (2025) offers an instructive discussion on what may be considered legitimate use of GenAI for literature review, or as the authors put it, "*what we should allow GenAI to do*" (p. 1). Akin to other research methods, the potential of GenAI misuse in submitted literature review papers raises challenging questions for editors and reviewers, who are confronted with rising submission numbers but lack effective means to detect the misuse of GenAI.

There is already a concerning "tendency to offload human cognition and intelligence to GenAI which can have potentially dysfunctional consequences that are, at this time, largely unknown" (Susarla et al., 2023: 405). We believe it is prudent to also seriously consider what GenAI means for broader scientific progress. While full consequences of using GenAI for literature reviews remain uncertain, quite likely, for some research teams, the level of innovation will spike following the use of GenAI for literature reviews, whereas for other teams, it may be to their detriment. This brings a research opportunity to understand when, and under what conditions, the negative or the positive tendency develops. Providing a comprehensive and contextualized answer to this question stands to benefit scientific progress and broadly human society, which depends on science and its development.

The utility of GenAI in propelling scientific inquiry, particularly in conducting literature reviews, significantly depends on the researchers' approach to integrating these technologies into their research efforts. Researchers endowed with a deep understanding of their investigative domains are aptly equipped to critically evaluate the outputs generated by tools such as ChatGPT and Gemini, aligning them with their domain knowledge to identify promising pathways to pursue in their research endeavors. Consequently, for seasoned scholars, GenAI holds the potential to substantially enhance research outcomes. Conversely, researchers with less prior exposure to the focal literature may encounter difficulties in accurately assessing the relevance and validity of GenAI-generated outputs, potentially leading to the exploration of less viable research avenues. In these instances, GenAI might inadvertently impede scientific progress even if the sheer volume of scientific papers grows. This may be another "the rich get richer, and the poor get poorer" scheme. This interplay underscores the indispensable role of domain-specific knowledge in maximizing the benefits of GenAI, highlighting the imperative for a synergistic integration of researcher acumen and technology to foster scientific advancement. The good news is, the debates about ways to synergize humans and AI are now abound, for example, in the context of future of human work, software development (Jain et al., 2021; Lukyanenko et al., 2025), the methodology of AI supported literature reviews can learn from and contribute to these debates.

## Concluding remarks

Current discussions on how GenAI could facilitate the conduct of literature reviews are characterized by the excitement of new opportunities, as well as cautious and critical commentaries. Building on the preceding commentary on AI-supported literature reviews (Wagner et al., 2022), we aim to develop a more substantive connection to the established methodological discourse, and offer a balanced view, suggesting for which tasks GenAI may be beneficial, and clarifying potential shortcomings. Currently, the design of research tools and services in this area is evolving rapidly, but effectively using GenAI to conduct review projects requires a particular set of skills, and a closer alignment with established methodological principles.

We hope this paper contributes to a constructive foundation for GenAI-supported literature reviews across science and in other settings (e.g., business, private), where making decisions based on prior literature is happening.

## ORCID iDs

Gerit Wagner https://orcid.org/0000-0003-3926-7717
Julian Prester https://orcid.org/0000-0003-2682-8036
Reza Mousavi https://orcid.org/0000-0002-1990-7767
Roman Lukyanenko https://orcid.org/0000-0001-8125-5918
Guy Paré https://orcid.org/0000-0001-7425-1994

## Funding

The authors received no financial support for the research, authorship, and/or publication of this article.

## Declaration of conflicting interests

The authors declared no potential conflicts of interest with respect to the research, authorship, and/or publication of this article.

## Supplemental material

Supplemental material for this article is available online.

## Notes

1. The approaches considered by Wagner et al. (2022) included popular AI techniques such as neural networks, random forests, decision trees, and pre-transformer natural language processing algorithms, such as LSA or LDA. These AI techniques remain popular and valuable for literature reviews, as outlined in the aforementioned paper. However, GenAI offers new opportunities and challenges not covered by and not relevant to these techniques (such as hallucinations and specific biases).
2. https://openai.com/index/dall-e-3/.
3. https://openai.com/sora/.
4. https://notebooklm.google/.
5. https://openai.com/index/hello-gpt-4o/.
6. Detailed explanations of LLM architectures and training paradigms are provided in the supplemental materials available online.
7. Please note that multimodal GenAI such as GPT-4o outperform text-only LLMs such as GPT-4 even in natively text-based tasks. Please refer to https://crfm.stanford.edu/helm/lite/latest/ for a benchmark of multimodal GenAI models and LLMs.
8. https://huggingface.co/tasks.
9. https://crfm.stanford.edu/helm/lite/latest/.
10. When using PDF documents in a prompt, it is important to check potential restrictions in context windows, and considering options like splitting data input or using paid APIs.
11. Additional detail on the evaluation is provided in in the supplementary online material. This paper does not aim to provide a comprehensive benchmarking system for evaluating GenAI's performance in conducting literature reviews. However, we share several useful evaluation mechanisms. We also encourage future research to develop robust benchmarking frameworks for this purpose, building on works like Jin et al. (2021), which focused on the medical domain.
12. https://www.paperdigest.org/.
13. https://scite.ai/.
14. https://www.elsevier.com/products/scopus/scopus-ai.
15. https://consensus.app/.
16. https://elicit.com.
17. Obtaining access to unpublished work is essential in meta-analyses to address the "file-drawer problem" and reduce publication bias arising from the underrepresentation of non-significant results. Unpublished studies are typically acquired through personal communication, mailing lists, or institutional repositories.
18. Here are but a few of the famous catchy titles: "To Err is Human: Building a Safer Health System" (Committee on Quality of Health Care in America, 2000), "Guns, Germs, and Steel: The Fates of Human Societies" (Diamond, 1997), or "The Black Swan: The Impact of the Highly Improbable" (Taleb, 2007).
19. Please note that the URL (https://www.tei-c.org/ns/1.0) is a namespace identifier and is not associated with a document that can be accessed through the browser.
20. https://fs-ise.github.io/gen-ai-lr-prompts/.

## Author biographies

**Gerit Wagner** is a Full Professor of Information Systems Education at Frankfurt School of Finance & Management. Prior to joining Frankfurt School, he served as an Assistant Professor at Otto-Friedrich-Universität Bamberg, following a postdoctoral fellowship at HEC Montréal. His research focuses on literature reviews and knowledge synthesis, the impact of research methods, digital health, and digital platforms for knowledge work. His work has been published in leading international journals, including the *Journal of Strategic Information Systems*, *Journal of Information Technology*, *European Journal of Information Systems*, *Journal of Medical Internet Research*, *Information & Management*, and *Decision Support Systems*. He is a member of the Association for Information Systems and regularly serves as a reviewer for leading Information Systems journals and international conferences.

**Julian Prester** is a Senior Lecturer (Associate Professor) of Business Information Systems at the University of Sydney. His research interests lie at the intersection of technology, work and organisations, focusing on various forms of digital work such as remote organising, digital nomadism and platform work. Using ethnographic and in-depth qualitative field research methods as well as computational approaches including natural language processing, his research has been published in journals such as Journal of Information Technology, The Journal of Strategic Information Systems, Decision Support Systems, and Information & Management. He received his PhD as a Scientia Scholar from UNSW Sydney, where he was awarded the Weinstock Memorial Prize for the most innovative Information Systems thesis.

**Reza Mousavi** is an associate professor at the McIntire School of Commerce, University of Virginia. His research interests lie at the intersection of artificial intelligence (AI) and business, focusing on the societal impacts and economics of AI, user-generated content, and healthcare information systems. Reza employs machine learning, deep learning, natural language processing (NLP), and econometrics to examine the inner workings and applications of technologies such as large language models (LLMs). His work has been published in premier journals, including *Information Systems Research*, *Journal of Marketing*, *Journal of Management Information Systems*, *Journal of the Association for Information Systems*, and *Political Research Quarterly*. Prior to his academic career, he served as Lead Data Scientist at State Farm Insurance Co. and advised leading consulting firms on AI initiatives.

**Roman Lukyanenko** is an associate professor at the McIntire School of Commerce, University of Virginia. His research interests include data management and research methods (validity and artificial intelligence in literature reviews). Roman's ideas, tools, and methods to improve data management and research practices have received major awards, including the INFORMS Design Science Award, the Governor General of Canada Gold Medal, and the Hebert A. Simon Design Science Award. Roman's research has been published in 120 conferences and 36 journals, including Nature, MIS Quarterly, Information Systems Research, and ACM Computing Surveys. His 2019 paper on the quality of crowdsourced data received the Best Paper Award at MIS Quarterly. Roman is a co-author of "Systems Analysis and Design: An Object-Oriented Approach with UML" textbook published by Wiley.

**Guy Paré** is a Professor of Information Technology, holds the Research Chair in Digital Health and is an Associate Director of the Health Research Hub at HEC Montréal. His research has been published in top-tier scientific journals and presented at numerous international and national conferences. Dr. Paré's expertise in digital health has made him a sought-after authority for organizations such as the World Health Organization, the Department of Health in France, Canada Health Infoway, and the Québec Ministry of Health and Social Services. In recognition of his significant contributions to academia, he was named Fellow of the Royal Society of Canada in 2012 and more recently, Fellow of the International Academy of Health Sciences Informatics in 2023 and of the Association for Information Systems in 2025.